# Impact of Electrode Density of States on Transport through Pyridine-Linked Single Molecule Junctions

Olgun Adak[1], Richard Korytár [2], Andrew Y. Joe[1], Ferdinand Evers[2], and Latha Venkataraman[1]

[1]Department of Applied Physics and Applied Mathematics, Columbia University, New York, NY,

[2]Karlsruhe Institute for Technology, Karlsruhe, Germany

AUTHOR EMAIL ADDRESS: lv2117@ columbia.edu, ferdinand.evers@physik.uni-regensburg.de

ABSTRACT: We study the impact of electrode band structure on transport through single-molecule junctions by measuring the conductance of pyridine-based molecules using Ag and Au electrodes. Our experiments are carried out using the scanning tunneling microscope based break-junction technique and are supported by density functional theory based calculations. We find from both experiments and calculations that the coupling of the dominant transport orbital to the metal is stronger for Au-based junctions when compared with Ag-based junctions. We attribute this difference to relativistic effects, which results in an enhanced density of d-states at the Fermi energy for Au compared with Ag. We further show that the alignment of the conducting orbital relative to the Fermi level does not follow the work function difference between two metals and is different for conjugated and saturated systems. We thus demonstrate that the details of the molecular level alignment and electronic coupling in metal-organic interfaces do not follow simple rules, but are rather the consequence of subtle local interactions.

Understanding the interplay between different phenomena that govern charge transport at metal-molecule interfaces is fundamentally important for creating functional organic electronic devices.[1-3] The charge transport properties of metal point contacts and metal-molecule interfaces are dictated by the band structure of the metal[4-11] and the electronic structure of the molecule,[12-16] along with many local effects arising from their interaction such as hybridization, dynamic and static charge screening, surface dipole formation.[17-20] Here, we study charge transport through pyridine-terminated systems with conjugated and saturated backbones. Specifically we compare the experimental and theoretical charge transport properties of 4,4'-vinylenedipyridine (**1**), 4,4'-bipyridine (**2**), and 4,4'-ethylenedipyridine (**3**) with Au and Ag electrodes to investigate how the band structure of the metal mediates the charge transport in single-molecule junctions. We find that the conductance of the conjugated molecules, **1** and **2**, are greatly reduced when bound to Ag electrodes compared to Au electrodes, while conductance of **3**, which has a saturated bridge, is not changed significantly. In order to understand the origin of these trends, we probe the energy level alignment of the conducting molecular orbital and its coupling strength to the electrodes in these systems through a combination of experiment and density functional theory (DFT) calculations. We find that all systems exhibit significantly enhanced molecular orbital coupling to Au compared to Ag due to an enhanced density of d-states of Au around and above the Fermi energy. In the conjugated systems, we show that the level alignment is similar with both metals; for the saturated system however, the lowest unoccupied molecular orbital (LUMO) which mediates the charge transfer is significantly closer to the Fermi level with Ag electrodes. In all systems, we show that the molecular orbital is better coupled to Au electrodes when compared with Ag. For the conjugated systems, this results in a smaller conductance for the Ag junctions while for the saturated system, the measured conductance values are similar because the reduction in molecular orbital coupling strength is compensated by the change in the alignment of the LUMO relative to the metal Fermi level.

In this work, single-molecule junctions are created using the scanning tunneling microscope-based break-junction technique (STM-BJ), where an Au/Ag tip is repeatedly moved in and out of contact with an Au/Ag substrate in a molecular solution (see Figure 1A for schematic and SI Figure S1 for details).[21] A metal point contact is first formed and once this is broken, a molecule can bridge the gap between the broken electrodes to form a single-molecule junction. Thousands of conductance versus displacement traces are obtained from these repeated measurements and compiled into logarithmically binned one-dimensional histograms[13] without data selection for each system (see Figure 1B, 1C and 1D). We observe a clear peak in the conductance histogram, which is a signature of single-molecule junction formation (see SI Figure S2 for two-dimensional conductance-displacement histograms). All molecules show a double-peak conductance signature with Au electrodes, as has been previously demonstrated for pyridine-terminated molecules.[22] The lower peak is attributed to a vertical junction geometry, while the higher peak is due to tilted junction geometry which enhances the coupling to the electrodes.[23] The absence of such a double-peak with silver electrodes indicates that there is likely no distinct transition from tilted to vertical geometry during elongation which could be due to the reduced molecule-electrode van der Waals interaction for silver.[24, 25] For **1** and **2**, we find that Au junctions have a higher conductance than Ag junctions, as shown in Figure 1B and 1C. Unlike **1** and **2**, **3** shows the same conductance with both metals (Figure 1D).

Charge transport through a molecule bonded to metal electrodes involves a coherent transmission of electrons/holes from one lead to the other through the molecular orbital that dominates transport. Often, the transmission probability can be described using a Lorentzian form with only two parameters: the alignment of the molecular orbital relative to the metal Fermi level ($E_{level}$) and the broadening of this orbital due to hybridization with the metal electrode ($\Gamma$).[26, 27]

$$T(E) = \frac{(\Gamma/2)^2}{(E-E_{level})^2+(\Gamma/2)^2} .$$

Depending on $E_{level}$ and $\Gamma$, the current-voltage characteristics can be highly non-linear;[28, 29] in such cases, they can be used to determine the junction transport parameters.[30] Here, we employ a new experimental technique to determine the transport parameters. We apply an AC voltage in addition to a DC voltage across the molecular junction and measure the current over a bandwidth larger than twice the AC voltage frequency. We use a modified break junction technique for these measurements where instead of pulling the tip continuously away from the substrate, the tip is held at fixed displacement after pulling the junction apart as illustrated in Figure 2A. The current through the junction is then measured using a 50 kHz bandwidth while applying 150 mV AC voltage at 22 kHz in addition to a 1V DC voltage. The currents at the first and second harmonic frequencies are obtained by looking at the frequency domain representation of the measured current using the discrete Fourier transform (Figure 2B).

To see what gives rise to the currents at the first and the second harmonic frequencies, we obtain the Taylor series expansion of the junction current at $V_{DC}$:

$$I = \left( I(V_{DC}) + \frac{V_{AC}^2}{4} \frac{d^2I}{dV^2}\bigg|_{V=V_{DC}} + \frac{V_{AC}^4}{64} \frac{d^4I}{dV^4}\bigg|_{V=V_{DC}} + ... \right) + \left( \frac{dI}{dV}\bigg|_{V=V_{DC}} V_{AC} + \frac{V_{AC}^3}{8} \frac{d^3I}{dV^3}\bigg|_{V_o} + ... \right) \text{Sin}(\omega t)$$
$$- \left( \frac{V_{AC}^2}{4} \frac{d^2I}{dV^2}\bigg|_{V=V_{DC}} + \frac{V_{AC}^4}{48} \frac{d^4I}{dV^4}\bigg|_{V=V_{DC}} + ... \right) \text{Cos}(2\omega t) + higher\ harmonics$$

Here, I is the current through the junction, $V_{DC}$ is DC voltage, $V_{AC}$ is the amplitude of the AC voltage, and $\omega_{AC}$ is the frequency of the AC voltage.

The terms in the first parenthesis correspond to the DC current, while the terms in the second and third parenthesis represent the currents at the first and the second harmonic frequencies respectively. Next, we obtain the analytic expressions for the terms in the second and the third parenthesis in terms of $E_{level}$ and $\Gamma$ using the single Lorentzian model. In this model, assuming that the voltage drop across the junction is symmetric and that the coupling on the left and right electrode are the same, the current through the junction can be written as:

$$I(V) = \frac{G_0}{e}\int_{-\infty}^{\infty} dE \frac{\Gamma^2/4}{(E-\mathrm{E}_{level})^2 + \Gamma^2/4}\left(f(E-eV/2) - f(E+eV/2)\right)$$

where f(E) is the Fermi–Dirac distribution. Since all energy scales are much higher than average thermal energy (= 0.025 eV), we can neglect finite temperature effects as justified in the SI, and can therefore simplify the expression for current to yield:

$$I(V) = \frac{1}{2e}\Gamma G_0 \tan^{-1}\left(\frac{2\left(\frac{eV}{2} - \mathrm{E}_{level}\right)}{\Gamma}\right) + \frac{1}{2e}\Gamma G_0 \tan^{-1}\left(\frac{2\left(\mathrm{E}_{level} + \frac{eV}{2}\right)}{\Gamma}\right)$$

By differentiating the expression above with respect to voltage, one can obtain the terms in the Taylor series expansion of the junction current as a function of $E_{level}$ and Γ as shown explicitly in the SI. With these expression, we solve a set of two non-linear equations with two unknowns using the currents at the first and the second harmonic frequencies to get the $E_{level}$ and Γ.

Using this technique, we first determine $E_{level}$ and Γ in molecule **1** and **2** with Au electrodes. These systems have been well studied experimentally and shown to have a transmission function that is well-approximated by a single-Lorentzian form allowing us to benchmark the AC measurement.[22, 23, 31, 32] For these systems, we do not see any evidence that the low conducting vertical geometry junctions sustain the entire fixed displacement section. This is not surprising since these low-conducting junctions form at the apex of the electrodes and there is no room for mechanical and thermal perturbations. We therefore focus our analysis on junctions in the tilted, high conducting geometry. We find that on average $E_{level}$ and Γ are 1.1 eV and 40 meV for **1** while 1.2 eV and 60 meV for **2** with Au electrodes. These values are in good agreement with previously reported values for these systems measured under zero external bias.[32] This is interesting because in our current measurements, these systems are driven out of equilibrium under the high applied DC bias. There are several effects that would alter the transport characteristics of a single molecule junction under high bias voltages. First, under an external field, the molecular energy levels can be Stark-shifted.[31] Second, the external electric field can polarize the molecule and change molecular coupling strengths.[33] Third, charging of molecular orbitals due to bias would modify the level alignment.[34] Fourth, molecular vibrational modes could get excited due to an increase in the local temperature or due to inelastic scattering of electrons; this would result in sharp features in the dG/dV spectrum.[35, 36]

To investigate these effects, we perform measurement at three different bias voltages to see if there is a change in the level alignment or coupling strength with the DC voltage. We determine $E_{level}$ and $\Gamma$ at 0.5 V, 0.75 V and 1.0 V DC bias voltages for 1 and 2 with Au electrodes (see SI). We see that $\Gamma$'s do not change with the applied bias voltage, while the $E_{level}$ increases by about 0.1 eV when going from 0.5 V to 1.0 V DC voltage (see SI Figure S3). Since the measured couplings are very similar, we conclude that the molecule is not strongly polarized under bias, and furthermore, we can also conclude that dG/dV is not altered and thus local heating effects are not important under these experimental conditions. The fact that we see a slight shift in the level alignment with bias voltage could indicate that (a) we are charging the molecule under bias, (b) there is a slight Stark shift or (c) the transmission deviates slightly from a single Lorentzian form. Here, we do not attempt to identify the primary source of this shift; we conclude that the effect of bias voltage on level alignment is small and the transmission characteristics measured at high bias reflects the zero bias transmission.

To further benchmark the experimental technique, we carry out IV measurements on **1** and **2** with Au and create two-dimensional IV histograms for the two systems as detailed in the SI.[37] Next, using the measured values of $E_{level}$ and $\Gamma$, we generate IV traces and obtain simulated IV histograms for two system. As shown in SI Figure S4, we see excellent agreement between measured and simulated IV histograms in two systems further justifying the use of this AC-based experimental technique. However, on average, junctions show a rectification ratio of 1.3 (as shown in SI Figure S5). This would imply that the coupling on the two sides of the junction could differ by 30% which contradicts one of the assumptions that the experimental technique relies on. In the SI, we provide a detailed analysis of the experimental technique when the assumption of equal couplings is relaxed to allow for differences in $\Gamma$ between the two sides (see SI Figure S6). We find that on average the extent of the asymmetry in coupling observed in these systems is not large enough to incur significant errors in estimated $\Gamma$ and $E_{level}$ parameters. This conclusion is already apparent from agreement between parameters measured under different bias voltages, and the excellent agreement between the IV histograms measured and generated using $E_{level}$ and $\Gamma$. However, we should add that this technique is not applicable in its current from to the systems with inherent asymmetries such as molecules that have different terminal linkers.[38]

Having established that our method yields accurate transport parameters, we next use it to probe $E_{level}$ and $\Gamma$ in molecule **1** with Ag electrodes as the transmission in this system can be well approximated by a single Lorentzian as will be shown further below. We find that $E_{level}$ and $\Gamma$ are 1.0 eV and 14 meV respectively (see Figure 3A and 3B). For **2** with Ag, we cannot determine $E_{level}$ and $\Gamma$ due to a signal-to-noise limitation of our instrument as the current at the second harmonic frequency is at or below the experimental noise floor. This is consistent with the fact that **2** with Ag exhibits even smaller conductance than **1** with Ag. We note that the difference in LUMO energies relative to the Fermi level between Au and Ag junctions for **1** is quite small (0.1 eV) compared to the work-function difference between the two metals, which is reported as ranging from 0.6-0.8 eV.[34]

To gain further understanding into charge transfer characteristic of these systems, we turn to DFT calculations of model junctions for all three molecules bound to both Au and Ag electrodes. Our first-principles calculations of the electronic transmission are based on a finite cluster approach that uses the PBE generalized gradient-corrected exchange-correlation functional[39] as implemented in the FHI-AIMS code.[40] The junction contacts are modeled with two Au or Ag pyramids containing 31 atoms each cut from a face-centered lattice (lattice parameter 2.9387 Å for Au and 2.9331 Å for Ag). The junction axis is along the crystalline (100) direction. The molecules are bound to the apex atoms on each pyramid as shown in Figures 4A-4C modeling a vertical geometry, which allows for a direct comparison of the three systems. Furthermore, identical junction structures are used here with different metals to focus on the effect of band structure on transport properties. The positions of H, C and N atoms were optimized with a variant of the Broyden-Fletcher-Shanno-Goldfarb algorithm[40], until the forces dropped below 0.01 eV/ Å. The electron transmission through the junction was calculated with a Green's function approach applied to the composite electrode-molecule system and a simplified embedding self-energy.[41, 42]

We show in Figure 4 the transmission for all three molecules bound to both Ag and Au electrodes. On a qualitative level we see that in all cases the conductance is dominated by the LUMO. Moreover, for molecule **1** and **2,** the transmissions are reasonably well approximated by single Lorentzians as visible from the fits overlaid, validating the single-Lorentzian assumption for the experimental method. For molecule **3**, the peak shape is more complicated because of the proximity of higher unoccupied molecular orbitals. We see that for all molecules the LUMO level broadening is larger with Au than with Ag. Specifically, the Lorentzian fits comparing Au and Ag transmissions indicate that the broadening is larger by about a factor of two for both molecules. This is surprising because the Ag-N and Au-N bonding lengths differ only by 2% and the total densities of states of Ag and Au near the Fermi energy differ by roughly 10%.

To understand the origin of this difference between Au and Ag electrodes, we look at the cluster eigenstates with energies close to the transmission resonance, focusing on junctions formed with **1** (Figure 5A). We see that on the molecule the eigenstates resemble the LUMO of the gas-phase. For the Au junction, the wavefunction on the apex atom has angular nodes which are characteristic of Au-$d_{yz}$. In addition, one can see presence of the Au-$p_y$ orbital because the lobes that point to the molecule are bigger. These orbitals have the right symmetry to couple to the LUMO. In the case of Ag, we also see four lobes reminiscent of an Ag-$d_{yz}$ orbital on the apex atom; however, these are rotated, which diminishes the coupling to the LUMO. These eigenstates indicate the importance of the d-states in hybridizing with the LUMO. In the next step, we look at the density of states of the apex orbitals involved in coupling. As shown in Figure 5B, we observe that the spectral density of the $d_{yz}$-states is three times larger in Au than in Ag while the $p_y$-states have very similar spectral density at the Fermi level. The enhanced presence of d-states in Au is due to the higher position of the d-band edge, reflecting known relativistic effects.[43, 44] To demonstrate the role of relativity, we perform non-relativistic DFT calculations for molecule **1**. We observe that the relativistic corrections increase the width of the LUMO by 48% with Au, whereas with Ag the width is enhanced only 10% (see SI Figure S7). Therefore, we conclude that Au gives rise to an enhanced level broadening due to the larger contribution of d-states at the Fermi level. We note here that past density-function theory calculations that have compared the transmission of junctions formed with **2** using Pt and Au electrodes have found a similar increase in coupling with Pt when compared with Au, which was also attributed to an enhanced density of d-states at the Fermi level.[6]

We cannot extend our theoretical analysis to the level alignment and conductances, because the former is inaccurate due to approximations inherent in the exchange-correlation functionals. Specifically, DFT based HOMO and LUMO are artificially close to the Fermi level due to self-interaction and polarization errors.[45-47] Therefore, one cannot take the transmission value at $E_F$ as a measure of conductance. However, both the polarization correction and the self-interaction error are molecule specific and thus one would expect these to be largely unaffected by the nature of the electrode material as long as the electrode is a metal. Therefore, we can compare the level alignment for a given molecule with different metal electrodes, while keeping in mind that the absolute level alignment is not correct. Our experimental analysis of molecule **1** has shown that the level positions differ only by 0.1 eV between Ag and Au and a similar small difference is obtained from the DFT calculations as well. This is much smaller than the work-function difference between Ag and Au. For **3**, we see that LUMO is 0.4 eV closer to the Fermi level with Ag compared to Au while the coupling for Ag is smaller than that of Au. Thus for **3,** the reduction in coupling compensates the change in the level alignment explaining the similar measured conductances with Au and Ag electrodes.

Comparing the experimental and calculation results presented here, we find that the molecule-metal coupling strength is largely controlled by metal density of states. However, unlike what one would expect for non-interacting systems,[48] the observed relative molecular level alignment between Ag and Au junctions for the three molecules studied here cannot be explained by simply considering the difference in the metal work function. There are several factors that alter level alignment when a molecule is brought into contact with a metal: charge screening by the metal brings molecular orbitals closer to the Fermi level,[49] while charge transfer from/to the molecule and any dipole formed the metal-molecule bond can alter alignment.[10, 11, 50, 51] Because of these local interactions, level alignment of a molecular system with different metals does not strictly follow the difference in work function. Our results show further that even molecular backbone play role in how the local interactions dictate the level alignment. We thus demonstrate that pyridine linked molecules couple poorly to Ag electrodes compared to Au electrodes due to the reduced density of d-states, while the energy level alignment is dictated in part by the molecular backbone. These findings shed light on the role of the electrode band structure and the local electrostatic effects in determining the charge transfer properties at metal-molecule interfaces. A detailed study of the electrode band structure is therefore equally important in understanding the electronic structure of organic constituents for designing better metal-organic electronic devices.

**Acknowledgements:** The experimental portion of this work was supported by the NSF under award DMR-1122594 and the Packard Foundation. R.K. and F.E. gratefully acknowledge the Steinbuch Centre for Computing (SCC) for providing computing time on the computers HC3 and UC1 at Karlsruhe Institute of Technology (KIT).

**Figure Captions:**

Figure 1. (A) Schematic of molecular junctions formed using the STM-BJ technique. (B-D) Logarithmically binned conductance histograms for measurements of Ag and Au junctions formed with molecules **1**, **2** and **3** respectively. Inset shows molecular structures.

Figure 2. (A) Representative conductance and displacement traces plotted against time for the modified break junction technique. An AC voltage is applied between 80 and 200 ms as indicated by the shaded grey region. (B) Inset: Current measured for the trace shown (A). The magnitude of the DC current when the AC voltage is applied is around $10^{-7}$ A. Main Panel: Frequency domain representation of the current measured while the AC voltage is applied. The first and second harmonic peaks at 22 kHz and 44 kHz with magnitude around $10^{-8}$A and $10^{-9}$A are clearly visible in the frequency domain. These values are used to determine the level molecular orbital alignment and coupling as detailed in the text.

Figure 3. Histograms of (A) the energy difference between the LUMO and the metal Fermi level ($E_{LUMO}$) and (B) the molecular orbital coupling for Ag and Au junctions with molecule **1** ($\Gamma$).

Figure 4. Calculated DFT based transmission curves for Au and Ag junctions formed with (A) **1**, (B) **2**, and (C) **3**. Dashed lines are Lorentzian fits to the data. Inset: Junction structure for each system shown with Au.

Figure 5. (A) Isosurface plots of transmitted state close to the resonance for molecule **1** junctions with Ag (top) and Au (bottom). (B) Density of states for $d_{yz}$ and $p_y$ orbitals in Au and Ag.

# Figure 1

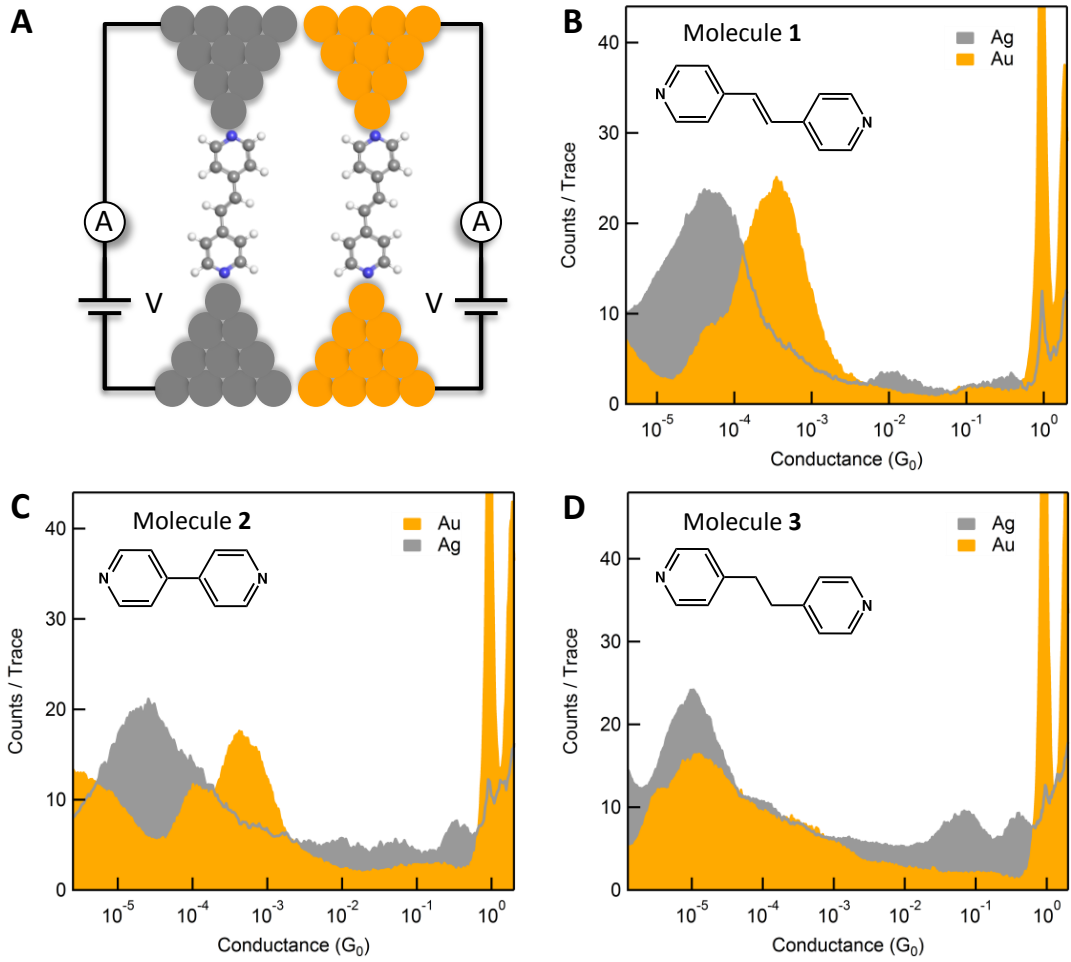

# Figure 2

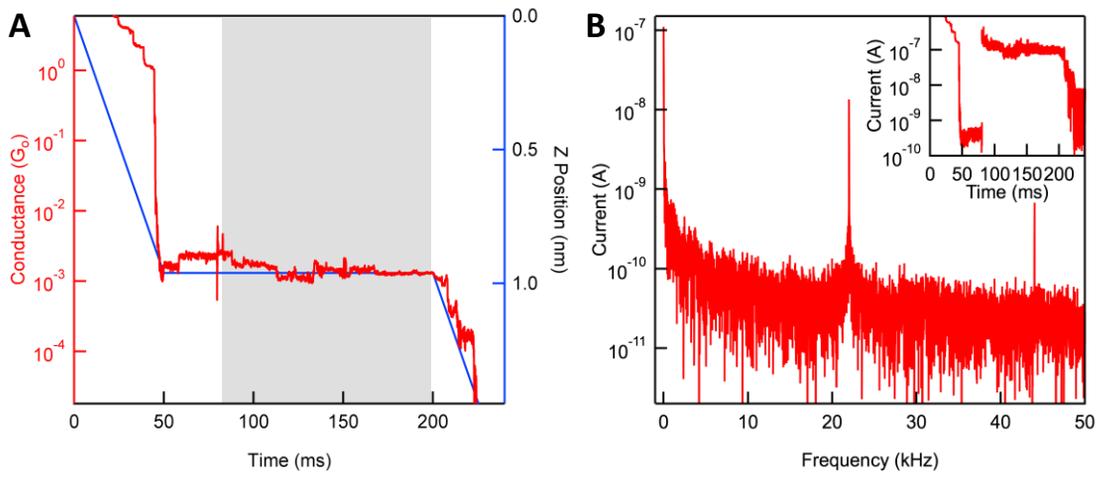

Figure 3

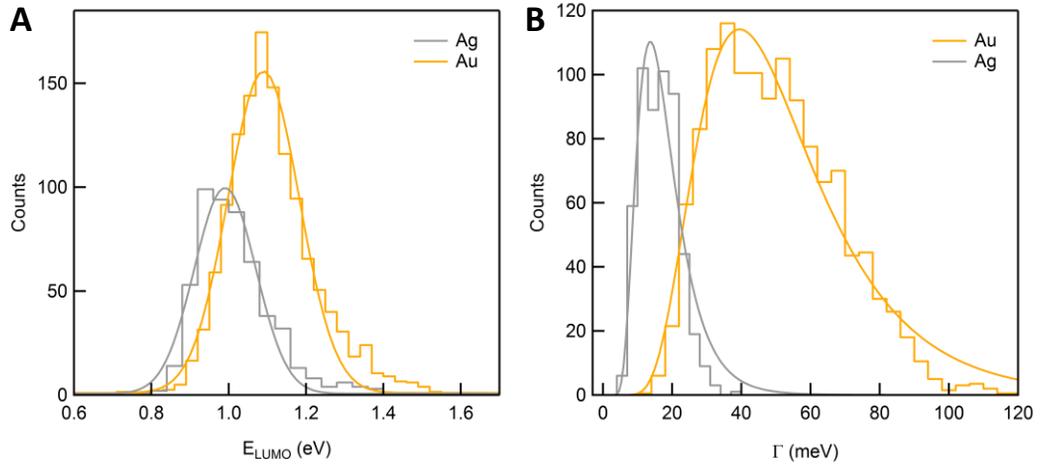

# Figure 4

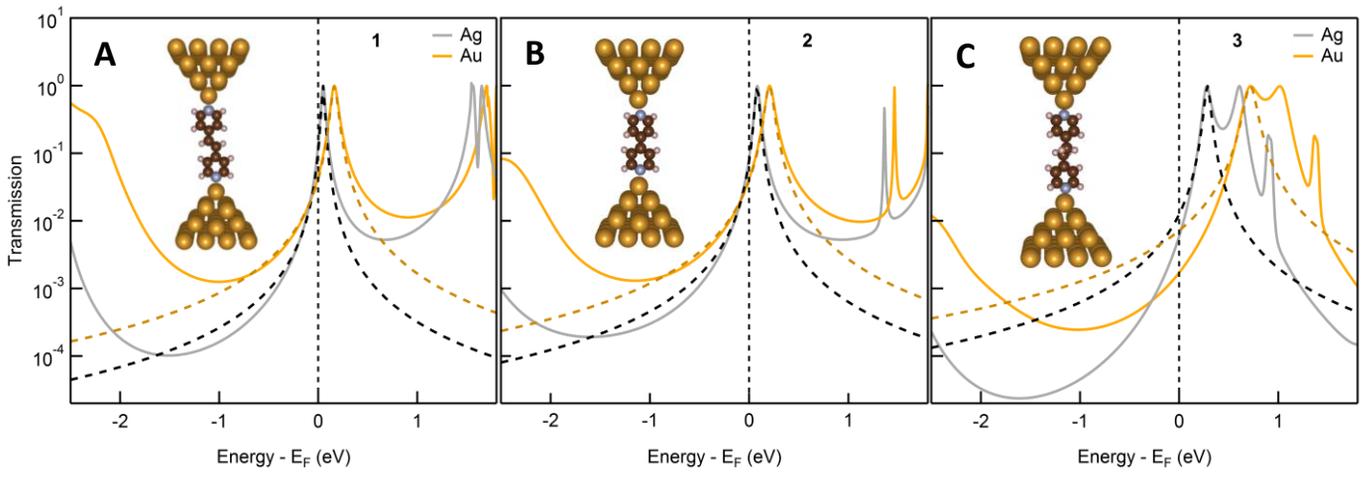

Figure 5

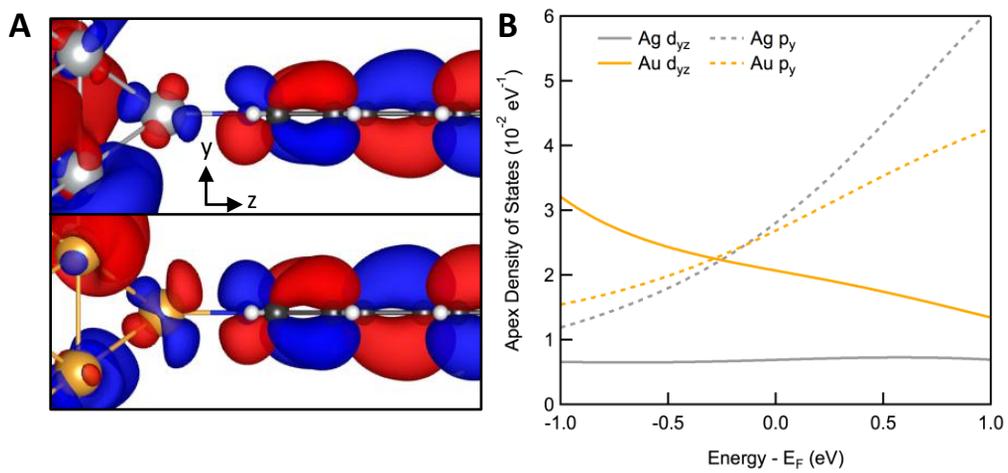